# Investigation of the mechanism of the anomalous Hall effects in Cr$_2$Te$_3$/(BiSb)$_2$(TeSe)$_3$ heterostructure

Seong Won Cho[1,2,†], In Hak Lee[3†], Youngwoong Lee[1,4], Sangheon Kim[1,5], Yeong Gwang Khim[6,7], Seung-Young Park[8], Younghun Jo[8], Junwoo Choi[3], Seungwu Han[2], Young Jun Chang[6,7], Suyoun Lee[1,9,*]

[1]Center for Neuromorphic engineering, Korea Institute of Science and Technology, Seoul 02792, Korea

[2]Department of Materials Science and Engineering, Seoul National University, Seoul 08826, Korea

[3]Center for Spintronics, Korea Institute of Science and Technology, Seoul 02792, Korea

[4]Department of Physics, Konkuk University, Seoul 05029, Korea

[5]Department of Materials Science and Engineering, Korea University, Seoul 02841, Korea

[6]Department of Physics, University of Seoul, Seoul 02504. Korea

[7]Department of Smart Cities, University of Seoul, Seoul 02504, Korea

[8]Center for Scientific Instrumentation, Korea Basic Science Institute, Daejeon 34133, Korea

[9]Division of Nano & Information Technology, Korea University of Science and Technology, Daejeon 34316, Korea

[†] These authors contributed equally.

* e-mail: slee_eels@kist.re.kr

**Abstract**

The interplay between ferromagnetism and the non-trivial topology has unveiled intriguing phases in the transport of charges and spins. For example, it is consistently observed the so-called topological Hall effect (THE) featuring a hump structure in the curve of the Hall resistance ($R_{xy}$) vs. a magnetic field ($H$) of a heterostructure consisting of a ferromagnet (FM) and a topological insulator (TI). The origin of the hump structure is still controversial between the topological Hall effect model and the multi-component anomalous Hall effect (AHE) model. In this work, we have investigated a heterostructure consisting of $Bi_xSb_{2-x}Te_ySe_{3-y}$ (BSTS) and $Cr_2Te_3$ (CT), which are well-known TI and two-dimensional FM, respectively. By using the so-called "minor-loop measurement", we have found that the hump structure observed in the CT/BSTS is more likely to originate from two AHE channels. Moreover, by analyzing the scaling behavior of each amplitude of two AHE with the longitudinal resistivities of CT and BSTS, we have found that one AHE is attributed to the extrinsic contribution of CT while the other is due to the intrinsic contribution of BSTS. It implies that the proximity-induced ferromagnetic layer inside BSTS serves as a source of the intrinsic AHE, resulting in the hump structure explained by the two AHE model.

# 1 Introduction

The interplay between the ferromagnetism and the non-trivial topology has unveiled intriguing phases in the transport of charges and spins, for example, the quantum anomalous Hall effect (QAHE) [1-5], the spin-orbit torque (SOT) [6-9], and the topological Hall effect (THE) [10-12]. In this regard, the heterostructure composed of a topological insulator (TI) and a ferromagnetic material (FM) not only provides a model system to study unexplored physics but also enables the development of novel energy-efficient spintronic devices [6,7,10,11,13,14]. Among the aforementioned intriguing phenomena, much attention has recently been paid to THE featuring a hump structure within the AHE hysteresis loop as it has been regarded as a fingerprint of skyrmions, the topologically-protected chiral spin textures [15]. Such an explanation is considered quite natural because of the strong spin-orbit coupling of a TI and the breaking of the inversion symmetry at the interface, which are the prerequisites of the Dzyaloshinskii-Moriya (DM) interaction to form skyrmions. Nevertheless, there was recently reported another explanation of the phenomenon [16-18]. After studying similar features observed in $SrTiO_3/SrRuO_3$ system, the authors proposed another viewpoint regarding those features as a result of competing contributions from two FM domains with distinguished coercive fields and the complementary sign of the AHE ("two-component AHE model" or "2AHE model") [16,17,19].

In this work, as a model system of the TI/FM heterostructure, we have investigated a heterostructure consisting of $Bi_xSb_{2-x}Te_ySe_{3-y}$ (BSTS) and $Cr_2Te_3$ (CT), which are well-known TI and layered ferromagnet with self-intercalation, respectively [20-27]. Both of them have the in-plane hexagonal symmetry and share the Te, which guarantees the formation of a high-quality heterostructure with a sharp interface avoiding unwanted effects due to intermixing and mechanical strain. It is found that the aforementioned hump structure appears only in the CT/BSTS heterostructure, not in the single film of CT and BSTS, indicating the critical role of the interface. Similar heterostructures, for example, some consisting of $Cr_2Te_3$ and $Bi_2Te_3$ [28-30] and others consisting of Cr-doped and V-doped $Sb_2Te_3$ [2,16,31], were examined to lead to interpretations different from each other for the observed hump structure. Here, we have tried two approaches to distinguish between the skyrmion-based model and the 2AHE-based model. The first is to use a TI (BSTS) with mostly suppressed bulk conductivity, which otherwise might lead to a misinterpretation. The second is to use the so-called "minor-loop measurement" [32], where the external magnetic field ($H$) is limited below the value at the extrema of the hump feature in one polarity of $H$ while it reaches over the saturation field in the other polarity during the measurement of the $R_{xy}$ vs. $H$ loop. It was reported that the shape of the hysteresis curve of the minor loop depended on the origin of the hump structure. In addition, we have performed a further analysis of the relationship between the Hall resistivity ($\rho_{xy}$) and the longitudinal resistivity ($\rho_{xx}$) of the CT/BSTS structure, which provides a clue about the origin of the phenomenon and underscores the interplay between a TI and an FM.

## 2 Results and Discussion

### 2. 1 Basic properties of the CT/BSTS heterostructure

Fig. 1a shows $\theta$-$2\theta$ x-ray diffraction (XRD) patterns of a single CT film (8 nm) and a CT/BSTS heterostructure (8 nm/100 nm) grown on an $Al_2O_3$(0001) substrate [21]. It shows that both of them are highly oriented along the c-axis and that BSTS is grown epitaxially on CT with keeping its own crystal structures despite the lattice mismatch of ~12% between BSTS and CT. It is attributed to both the van der Waals bonding nature of the BSTS and the low growth temperature of the top BSTS layer which is ~100 K lower than that of the bottom CT layer.

Fig. 1b shows the sheet resistance ($R_{sh}$) of an 8-nm thick CT film, a 100-nm thick BSTS film, and a CT(8 nm)/BSTS(100 nm) heterostructure as a function of temperature ($T$). The BSTS film shows an insulating behavior down to ~ 50 K, below which its $R_{sh}$ starts to saturate, consistent with the behavior of TI. $R_{sh}(T)$ of the CT film shows a change in the slope around the Curie temperature of $T_c$ = 170 K corresponding to the ferromagnetic transition [21]. In contrast, $R_{sh}(T)$ of the CT/BSTS heterostructure is observed to be less sensitive to temperature showing a broad peak around 195 K, which is due to the change of the dominating transport channel from the BSTS layer to the CT layer. Looking at the curve more closely (Fig. 1c), a change in the slope is observed around 76 K, which seems to indicate the existence of another ferromagnetic layer. It might be attributed to a proximity-induced ferromagnetic layer inside the BSTS layer or a Cr-doped BSTS layer possibly formed by the intermixing between the BSTS and CT layers [19,33-35]. This will be further discussed in the later part.

Fig. 1d shows the field-cooled magnetization of the CT/BSTS film as a function of temperature with the magnetic field (100 Oe) applied parallel ($H_{para}$) and perpendicular ($H_{perp}$) to the film plane, respectively. From the temperature-dependent magnetization curve, clear ferromagnetic transition is observed under the 170 K with the out-of-plane (OOP) direction consistent $R_{sh}(T)$ of the CT shown in Fig. 1b while the in-plane (IP) direction doesn't show an increment of magnetization. Fig. 1e shows the magnetization of the CT/BSTS film as a function of $H$ at 70 K under both $H_{para}$ and $H_{perp}$ configurations, showing a large hysteresis in the OOP direction with the coercive field of ~0.8 T. On the other hand, the in-plane (IP) magnetization shows the non-saturating behavior up to ± 2T indicating the strong perpendicular magnetic anisotropy energy of the CT [36,37]. The magnetic properties such as $T_c$ and strong perpendicular magnetic anisotropy of CT/BSTS are consistent with the previously reported $Cr_2Te_3$ thin films [14,36,38].

**2. 2 Anomalies in the Hall effect of the CT/BSTS heterostructure**

Fig. 2 shows the Hall resistivity trace ($\rho_{xy}(H)$) of BSTS, CT, and CT/BSTS in the temperature range of 50 ~ 150 K. For the single BSTS film (Fig. 2a), $\rho_{xy}(H)$ is observed to be linear resulting in the estimations of the carrier density ($n$ = 4.7x10$^{18}$ /cm$^3$) and the mobility ($\mu$ = 105 cm$^2$/Vs) at 50 K. In addition, it is observed that the slope of $\rho_{xy}(H)$ increases with lowering the temperature. Together with $R_{sh}(T)$ of the BSTS film, it implies that the bulk conductivity of the BSTS is sufficiently suppressed at relatively low temperatures (T < 100K) to make the topologically-protected surface channel dominate the carrier transport in the BSTS layer. Meanwhile, as shown in Fig. 2b, $\rho_{xy}(H)$ of the single CT film shows a counter-clockwise hysteresis loop whose coercive field increases with lowering the temperature. The amplitude of the AHE decreases with lowering the temperature, which is not common behavior in a magnetic film. Since the anomalous Hall resistance is proportional to the magnetic moment of the film (e. g., $R_{AHE} \propto R_s M$, $R_s$ is the anomalous Hall coefficient and $M$ is the magnetic moment), an anomalous Hall resistance should increase until the magnetic moment reaches the saturation magnetic moment as temperature decreases. However, the magnetization of CT/BSTS has almost saturated under 100 K (Fig. 1d), and the anomalous Hall resistance is limited by $R_s$ which is determined by both the mechanism of AHE and the longitudinal resistivity [12,38-40]. So, the decrease of $\rho_{xy}$ of CT could be accounted for both the decrease in $R_{sh}$ of the CT film (Fig. 1b) and the extrinsic AHE mechanism [41].

On the other hand, in Fig. 2c, the CT/BSTS heterostructure shows intriguing features such as the sign-reversal of the AHE at 50 K (from the counter-clockwise in CT to the clockwise in CT/BSTS) and a hump structure around $H$ = 1 T which is the similar value of the coercive field clearly observed at 50 K and 70 K, the aforementioned THE-like feature.

The measured $\rho_{xy}(H)$ in Fig. 2c can be decomposed into $\rho_{xy}(H) = \rho_{OHE}(H) + \rho_{AHE}(H) + \rho_{hump}(H) = \rho_{OHE}(H) + \rho'_{xy}(H)$, where the three terms in the middle represent the ordinary Hall resistivity, the anomalous Hall resistivity, and the hump-related resistivity in order. To focus on the THE-like feature, we have subtracted $\rho_{OHE}(H)$ from $\rho_{xy}(H)$ by assuming that $\rho_{OHE}(H)$ is not related to the spontaneous magnetization and the hump-related mechanism, that is to say, $\rho_{OHE}(H)$ should cross the origin (for extracting the nonlinear $\rho_{OHE}(H)$ from the measured $\rho_{xy}(H)$, we used the two-band model fitting method. For the details, see Fig. S2 in the Supplementary Information). $\rho'_{xy}(H)$ of the CT/BSTS heterostructure is plotted at temperatures from 50 K to 80 K as shown in Fig. 3a. Note that the hump structure is the most conspicuous at 60 K with $\rho_{hump}(H)$ overwhelming $\rho_{AHE}(H)$. Furthermore, at the same temperature, it is found that the sign of the AHE is reversed from the clockwise direction to the counter-clockwise direction. Above 60 K, the hump structure decays slowly, completely disappearing above 80 K. Interestingly, this temperature is near 76 K where the longitudinal resistance of the CT/BSTS structure shows a change in slope as shown in Fig. 1c. The amplitude and the coercivity of $\rho'_{xy}(H)$ in the temperature range of 50 ~ 150 K are summarized in Fig. 3b and 3c, respectively. It is observed that $H_c(T)$ can be described by Kneller's law [42] as $H_c = H_0\,(1 - T/T_B)^\alpha$ with $\alpha$ = 1/2, implying the single magnetic domain structure [43] (Fig. 3d).

Fig. 3e shows the minor loops of $\rho'_{xy}(H)$ of the CT/BSTS heterostructure at 60 K with varying negative bound ($H^n_{max}$) of $H$ and keeping the positive bound ($H^p_{max}$) at 3 T. In previous studies [17,32], it was reported that the minor-loop has an appearance depending on the mechanism of the hump structure, the skyrmion-based or the 2AHE-based model. When $H^n_{max}$ = -1 T, slightly higher than the field at the peak of the hump (= $-H_{hump}$), it is clearly observed that the minor loop of $\rho'_{xy}(H)$ forms a square-shaped hysteresis curve. Furthermore, in that case, $\rho'_{xy}(H)$ does not show a dip structure at $H = H_{hump}$, which is a conjugate to the hump feature at $H = -H_{hump}$ and appears in the full-loop measurement. Finally, when $H^n_{max}$ = -1.5 T slightly higher than the magnetic field at the saturated magnetization($-H_{sat}$ ($\approx$ -2 T)), $\rho'_{xy}(H)$ curve for the $H$-sweep in the positive direction is offset from the full-loop measurement, resulting in the size of the dip at $H = -H_{hump}$ smaller than that in the full-loop measurement. All these observations, Fig. 1c and Fig. 3e, are consistent with the sum of hysteresis loops of ferromagnetic domains or layers, supporting the 2AHE-based model for the hump structure in our CT/BSTS heterostructure.

## 2. 3 Analyses based on the two-channel AHE model

Based on the 2AHE model for the hump structure observed in the CT/BSTS structure, we have further analyzed the properties of the second ferromagnetic layer other than the CT layer. As mentioned in the discussion of the result in Fig. 1c, the layer might be attributed to a proximity-induced ferromagnetic layer inside the BSTS layer or a Cr-doped BSTS layer. To differentiate each component in the two-channel AHE, we have fitted the measured curve to an approximate form given by the sum of two AHE components, each of which is empirically described by the hyperbolic tangent function.

$$\rho_{AHE}^{tot}(H) = \rho_{AHE}^{neg}(H) + \rho_{AHE}^{pos}(H)$$
$$= -R^{neg} \tanh[\omega^{neg}(H - H_c^{neg})] + R^{pos} \tanh[\omega^{pos}(H - H_c^{pos})] \quad \ldots (1)$$

Here, the expressions of "*neg*" and "*pos*" as the superscript are used for representing the negative and the positive AHE, respectively, with the positive meaning the counter-clockwise AHE as observed for the CT single film as shown in Fig. 2b. $R^{neg}$, $R^{pos}$, $\omega^{neg}$, $\omega^{pos}$, $H_c^{neg}$, and $H_c^{pos}$ are all positive constants as fitting parameters.

Fig. 4a shows the result of the curve fitting at 50 K as a representative example, releasing $\rho_{AHE}^{pos}(H)$ and $\rho_{AHE}^{neg}(H)$. Repeating the same at various temperatures, we have obtained the temperature dependence of each AHE component as shown in Fig. 4b and 4c (for the results of the curve fitting at various temperatures, see Fig. S3). The fitting parameters $H_c^i$ and $R^i$ (*i*=pos, neg) are plotted in Fig. 4d and 4e as a function of temperature, respectively. Note that $R^{neg}$ and $R^{pos}$ show the temperature dependence opposite to each other while both $H_c^{neg}$ and $H_c^{pos}$ decrease with increasing temperature. In addition, note that the temperature dependence of $\rho_{AHE}^{pos}(H)$ resembles that of the CT single film as shown in Fig. 2b. Therefore, we believe that $\rho_{AHE}^{pos}(H)$ is attributed to the CT layer. On the other hand, considering that $\rho_{AHE}^{neg}(H)$ decreases with increasing temperature, we believe that $\rho_{AHE}^{neg}(H)$ is associated with the BSTS layer whose longitudinal resistivity decreases with increasing temperature. Therefore, in Fig. 4f and 4g, we have plotted $R^{pos}$ and $R^{neg}$ as a function of the longitudinal resistivities ($\rho_{CT}$ and $\rho_{BSTS}$) of the CT film and the BSTS film, respectively. Indeed, it is observed that $R^{pos}$ is linearly proportional to $\rho_{CT}$ while $R^{neg}$ shows a superlinear dependence on $\rho_{BSTS}$ ($R^{neg} \sim (\rho_{BSTS})^{1.8}$).

It is theoretically known that $\rho_{AHE}$ is proportional to the longitudinal resistivity, $\rho_0$, with being described by a power law [41], $\rho_{AHE} \sim (\rho_0)^\beta$. Here, the exponent *β* depends on the mechanism of the AHE. *β*=1 corresponds to the extrinsic skew scattering mechanism. *β* = 2 represents the AHE resulting from the intrinsic mechanism related to Berry's phase which is only affected by the electronic band structure of the material. Therefore, the obtained result of *β* = 1.8 seems to support the intrinsic picture that the negative AHE component might be the proximity-induced ferromagnetic layer inside BSTS which has a nonzero Berry's phase [9,44-47].

## 3 Conclusions

In a previous report[15], the authors investigated the $R_{xy}(H)$ loop of heterostructures consisting of (top) $Cr_2Te_3/Bi_2Te_3$ (bottom), $Bi_2Te_3/Cr_2Te_3$, and $Cr_2Te_3/Bi_2Te_3/Cr_2Te_3$. The hump structure was observed in $Cr_2Te_3/Bi_2Te_3$ and $Cr_2Te_3/Bi_2Te_3/Cr_2Te_3$ structures, but not in the $Bi_2Te_3/Cr_2Te_3$ structure. The authors interpreted the observed hump structure as a result of scatterings by the skyrmions without an explanation of the absence of such a feature in the $Bi_2Te_3/Cr_2Te_3$ structure. In another previous work[17], the authors investigated the $R_{xy}(H)$ loop of heterostructures consisting of (top) Cr-doped $Sb_2Te_3/Sb_2Te_3$ (bottom), $Sb_2Te_3$/V-doped $Sb_2Te_3$, and Cr-doped $Sb_2Te_3/Sb_2Te_3$/V-doped $Sb_2Te_3$. The authors showed that the sign of the AHE in the $Sb_2Te_3$/V-doped $Sb_2Te_3$ could be controlled by the thickness of $Sb_2Te_3$ and by the gate voltage applied to the structure. In addition, in the $R_{xy}(H)$ loop of the tri-layer structure, they observed the hump structure which was interpreted by the 2AHE model.

In this work, we have performed an in-depth study on the $R_{xy}(H)$ loop of the BSTS/CT heterostructure. We have observed the hump structure in the BSTS/CT heterostructure, but not in the single CT film. In addition, from the minor-loop measurement, the $R_{xy}(H)$ loop has been found to show behavior consistent with the expectation of the 2AHE model. Furthermore, by analyzing the temperature dependence of the $R_{xy}(H)$ loop of the heterostructure, we have found that the anomalous Hall resistivity ($\rho_{AHE}$) in one channel depends on its longitudinal resistivity ($\rho_{xx}$) with being described by a power law, $\rho_{AHE} \propto \rho_{xx}^{1.8}$. It reminds us of the intrinsic AHE mechanism by nonzero Berry's phase in that AHE channel, implying that it might be a proximity-induced ferromagnetic layer inside BSTS. These results suggest that the in-depth magneto-transport analysis may be applicable to other non-trivial AHE phenomena.

An intriguing observation is that the hump structure survives up to at least 80 K (see Fig. S2 in the Supplementary Information). This temperature is consistent with the temperature (~76 K) at the ferromagnetic transition of the interfacial layer in the BSTS/CT structure conjectured from the temperature dependence of the $R_{sh}$ of the CT/BSTS structure (Fig. 1c). Therefore, it implies that a ferromagnetic topological insulator with high $T_c$ can be induced by the proximity effect of an adjacent ferromagnet, which raises the possibility to study the unexplored physics of the ferromagnetic topological insulator at an increased temperature and further to realize novel energy-efficient spintronics-based electronic devices [27].

**4 Methods**

**4. 1 Film growth and characterization**

The Cr$_2$Te$_3$ (CT) thin film was grown on an Al$_2$O$_3$ (0001) substrate at a growth temperature of 350 ℃. Co-evaporation of Cr (99.995%) and Te (99.999%) was carried out at the base pressure = ~5x10$^{-8}$ Torr in a vacuum chamber equipped with an electron-beam (E-beam) gun and effusion cells. Details of the growth process of CT film can be seen in the previous report [21].

Then, Bi$_x$Sb$_{2-x}$Te$_y$Se$_{3-y}$ (BSTS) was grown ex-situ in another thermal evaporator. The BSTS was grown by thermal co-evaporation of Bi, Sb, Te, and Se atomic sources in a chamber with a base pressure of ~3x10$^{-8}$ Torr. Growth was carried out at a substrate temperature of 200 ℃ to a thickness of 100 nm. The composition of BSTS was optimized as $(Bi_{0.47}Sb_{0.53})_2(Se_{0.44}Te_{0.56})_3$ in the previous study to have maximally suppressed bulk conduction [22].

Before the growth of the BSTS layer, half of the CT film was masked by a sapphire substrate to release a specimen with half consisting of only the CT layer and the other half of CT/BSTS. In this way, possible errors were minimized by avoiding the effect of the variations of samples on the transport properties.

Structural analysis of the grown thin film was performed through an X-ray diffractometer (ATX-G, Rigaku). The magnetization measurements were carried out using a superconducting quantum interference device magnetometer (SQUID-VSM, Quantum Design Inc.) Two modes (zero-field cooling and field cooling) were used for temperature-dependent magnetization measurements with a fixed magnetic field of 100 Oe.

## 4. 2 Carrier transport measurement

The longitudinal resistance ($R_{xx}$) and the transverse (or Hall) resistance ($R_{xy}$) were measured by the conventional van der Pauw method. The samples of CT and CT/BSTS films were cut to a square of the size of 0.5x0.5 cm$^2$ and electrical contacts were made at the four corners by the indium press method. After wiring the electrical contacts, the samples were loaded into a commercial cryogen-free cryostat (Cmag Vari.9, Cryomagnetics Inc.). For the measurement, a Source-Measure Unit (2612A, Keithley Inc.) as a current source and a nano-voltmeter (2182, Keithley Inc.) as a voltage meter were used.

**Supplementary Information**

The online version contains supplementary material available at

**Fig. S1.** Magnetoresistance vs. H curves of BSTS, CT, and CT/BSTS. **Fig. S2.** Scheme of decomposing the Hall resistivity. **Fig. S3.** Separation of two AHE components at 50 – 90 K

**Acknowledgments**

Not applicable.

**Author contributions**

SWC and IHL contributed equally to the work. SL planned the experiment and supervised the research. SWC and YL conducted the characterizations of the transport properties. SH and YJC advised important ideas for analyzing transport data. SWC and SK grew and characterized the BSTS films. IHL and YGK grew and characterized the CT films. SYP, YJ, and JC conducted the magnetic measurement. All the authors took part in the writing of the manuscript.

**Funding**

This work was supported by the Korea Institute of Science and Technology (KIST) through 2E31550 and by the National Research Foundation program through NRF-2021M3F3A2A03017782, 2021M3F3A2A01037814, 2021M3F3A2A01037738, and 2021R1A2C3011450. 2020R1A2C200373211, [Innovative Talent Education Program for Smart City] by MOLIT.

**Availability of data and materials**

Not applicable.

**Declarations**

**Competing interests**

The authors declare no competing financial interest.

**Author details**

[1]Center for Neuromorphic engineering, Korea Institute of Science and Technology, Seoul 02792, Korea. [2]Department of Materials Science and Engineering, Seoul National University, Seoul 08826, Korea. [3]Center for Spintronics, Korea Institute of Science and Technology, Seoul 02792, Korea. [4]Department of Physics, Konkuk University, Seoul 05029, Korea. [5]Department of Materials Science and Engineering, Korea University, Seoul 02841, Korea. [6]Department of Physics, University of Seoul, Seoul 02504. Korea. [7]Department of Smart Cities, University of Seoul, Seoul 02504, Korea. [8]Center for Scientific Instrumentation, Korea Basic Science Institute, Daejeon 34133, Korea. [9]Division of Nano & Information Technology, Korea University of Science and Technology, Daejeon 34316, Korea

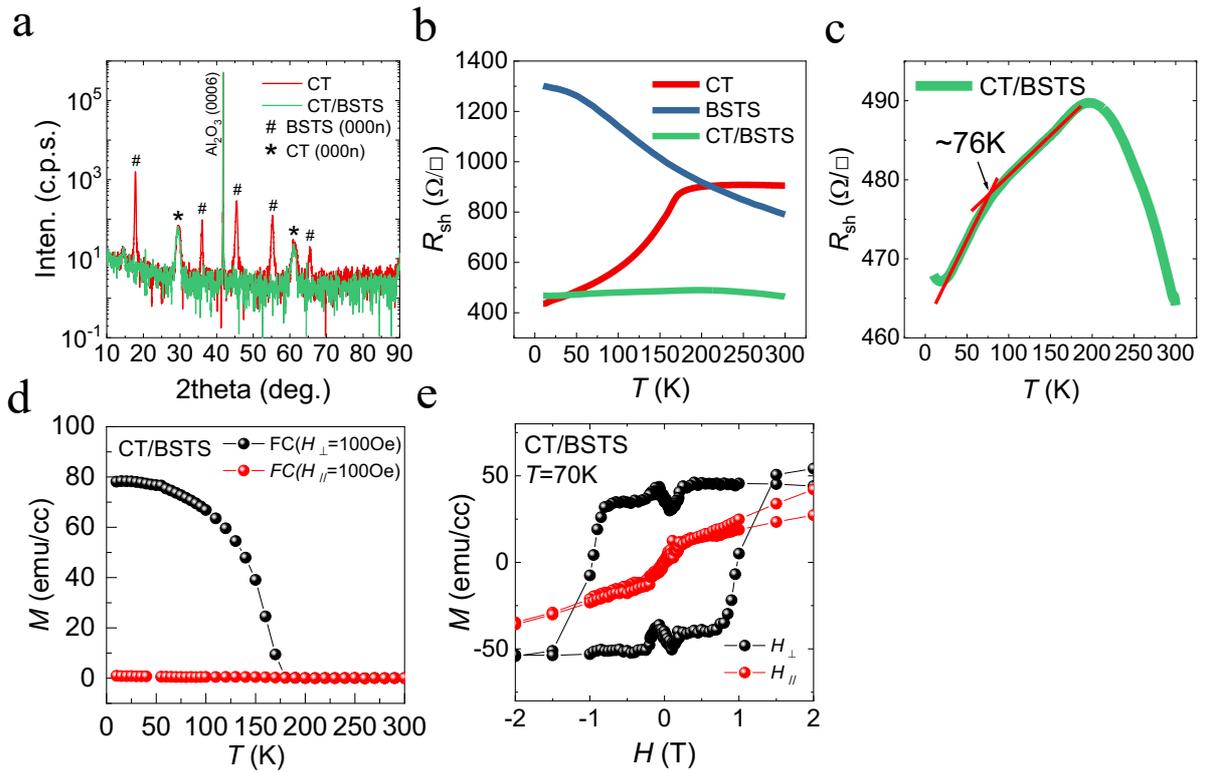

**Fig. 1** Basic physical properties of CT and BSTS. **a** XRD theta-2theta scans of CT (red) and CT/BSTS (green). **b** Sheet resistance as a function of temperature $R_{sh}(T)$ of CT (red), BSTS (blue), and CT/BSTS heterostructure (green). **c** An expansion of $R_{sh}(T)$ of CT/BSTS to highlight the bump at ~76 K. **d, e** The magnetization of CT/BSTS heterostructure as a function of temperature (**d**) and external magnetic field (**e**) at $T$ = 70 K with the in-plane (red) and the out-of-plane (black) magnetic field directions. All the measurements here use the 8 nm CT and 100 nm BSTS films.

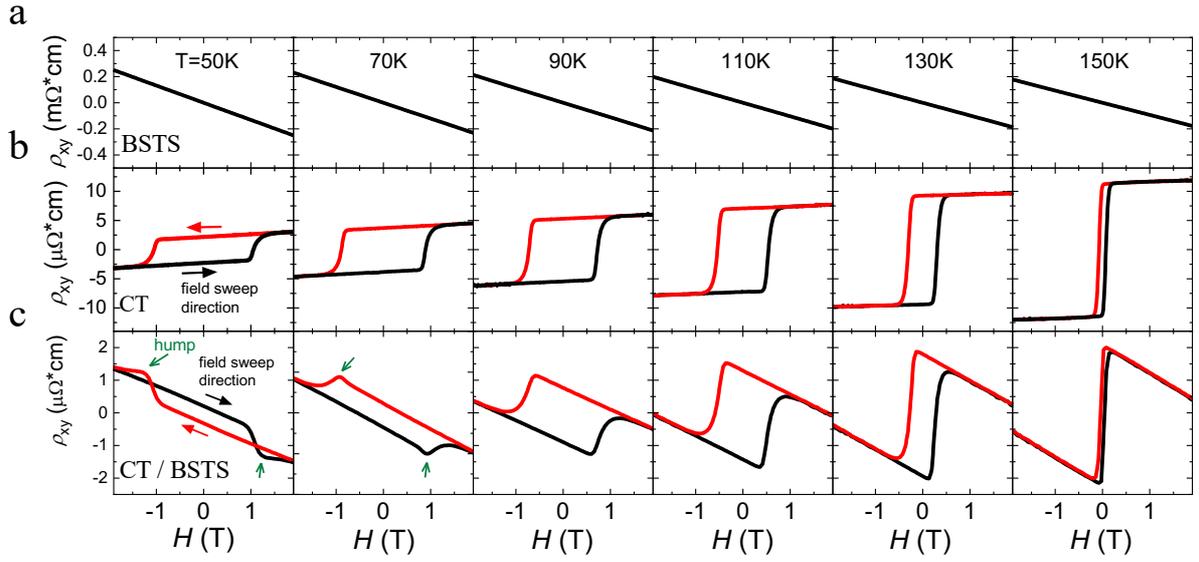

**Fig. 2** The evolution of Hall resistivity trace ($R_{xy}(H)$) of a CT/BSTS heterostructure. **a, b, c** The Hall resistivity as a function of the magnetic field of BSTS (**a**), CT (**b**), and CT/BSTS (**c**), respectively. The black and red curves represent the cases for ascending and descending magnetic field sweeps as indicated by black and red arrows, respectively. The appearance of the hump structure is highlighted by the green arrows in (**c**). The measurement temperatures (50 – 150 K) are indicated at (**a**).

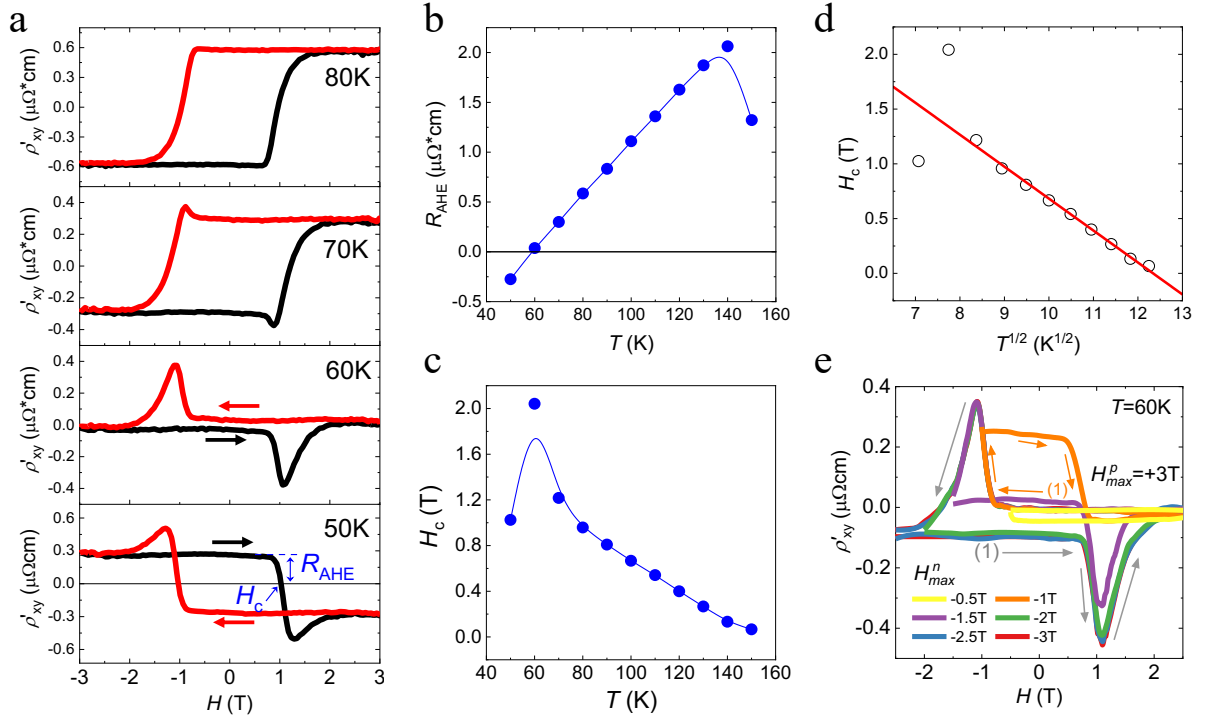

**Fig. 3** Analysis of the hump structure in the CT/BSTS heterostructure. **a** The reduced Hall resistivity trace ($\rho'_{xy}(H) = \rho_{AHE} + \rho_{hump}$) of the CT/BSTS heterostructure at temperatures of 50, 60, 70, and 80 K. **b, c** The amplitude of anomalous Hall resistivity, $R_{AHE}$ (**b**) and the coercive field, $H_c$ (**c**) as a function of temperature. The definitions of $R_{AHE}$ and $H_c$ are shown in panel **a**. **d** The Kneller's plot of $H_c$ vs. $T^{1/2}$. **e** The minor-loop measurement at 60 K with increasing $H^n_{max}$ = -0.5 – -3 T and $H^p_{max}$ = 3 T.

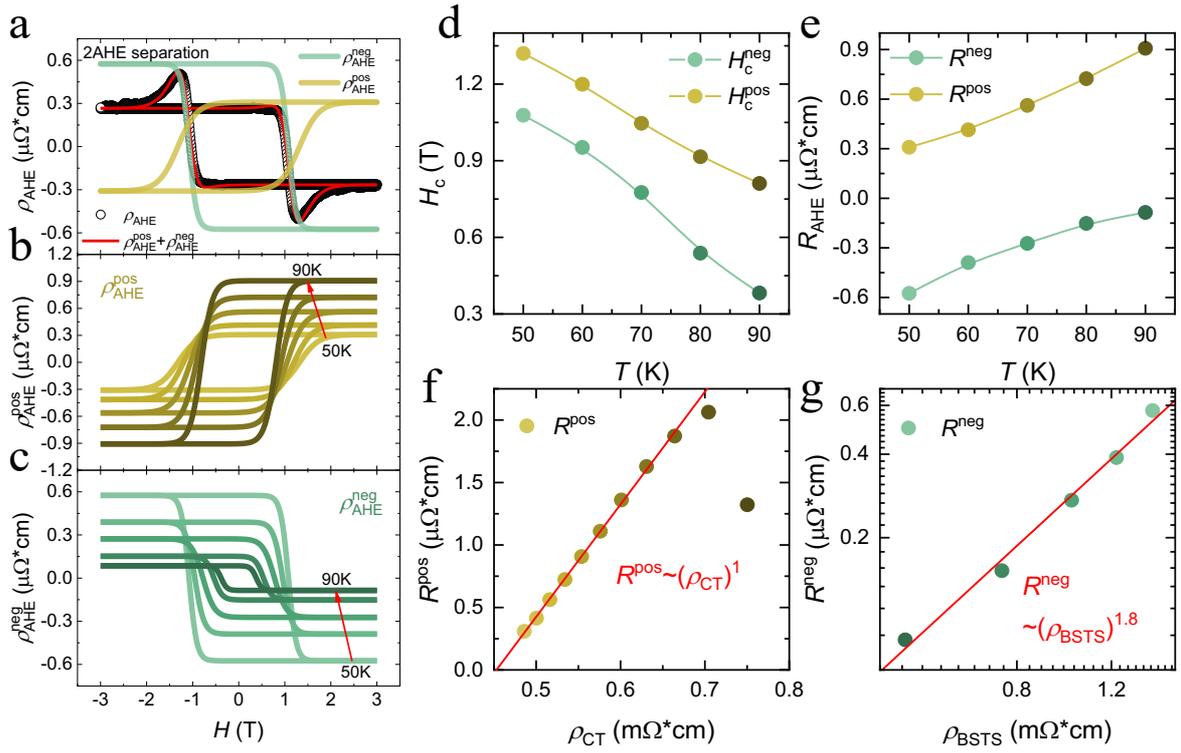

**Fig. 4** 2AHE model analysis of the hump structure of the CT/BSTS heterostructure. **a** Separation of $\rho_{AHE}(H)$ into the $\rho_{AHE}^{pos}$ and $\rho_{AHE}^{neg}$ at T=50 K. **b, c** $\rho_{AHE}^{pos}$ (b) and $\rho_{AHE}^{neg}$ (c) with varying the temperature in the range of 50 – 90 K. **d, e** The coercive fields ($H_c^{pos\,(neg)}$) (d) and the amplitude of anomalous Hall resistivities ($R_{pos\,(neg)}$) (e) of the negative (green) and the positive (yellow) AHE as a function of temperature. **f** Plot of $R^{pos}$ as a function $\rho_{CT}$, showing the linear dependence of $R^{pos} \propto \rho_{CT}$. **g** Log-log plot of $R^{neg}$ vs. $\rho_{BSTS}$, showing the power law dependence of $R^{neg} \propto \rho_{BSTS})^{1.8}$.